\documentclass[twocolumn,american,english,aps,prb,showkeys,amsmath,amssymb,superscriptaddress,a4]{revtex4-1}
\usepackage[T1]{fontenc}
\usepackage[latin9]{inputenc}
\usepackage{babel}
\usepackage{amsmath}
\usepackage{amssymb}
\usepackage{graphicx}
\usepackage{subscript}
\usepackage[unicode=true]
 {hyperref}
\usepackage{breakurl}

\makeatletter

\newcommand{\lyxmathsym}[1]{\ifmmode\begingroup\def\b@ld{bold}
  \text{\ifx\math@version\b@ld\bfseries\fi#1}\endgroup\else#1\fi}

\providecommand{\tabularnewline}{\\}

\usepackage{multirow}\usepackage{bm}\usepackage{gensymb}\usepackage{threeparttable}

\makeatother

\begin{document}

\title{Electron and phonon interactions and transport in ultra-high-temperature
ceramic ZrC}
\selectlanguage{american}%

\author{Thomas A. Mellan}
\email{t.mellan@imperial.ac.uk}

\selectlanguage{american}%

\affiliation{Thomas Young Centre for the Theory and Simulation of Materials, Department
of Materials, Imperial College London, Exhibition Road, London SW7
2AZ, UK}

\author{Alex Aziz }

\affiliation{Department of Chemistry, University of Reading, Whiteknights, Reading
RG6 6AD, United Kingdom}
\selectlanguage{english}%

\author{Yi Xia}
\selectlanguage{american}%

\affiliation{Department of Materials Science and Engineering, Northwestern University,
Evanston, IL 60208, USA}
\selectlanguage{english}%

\author{Ricardo Grau-Crespo}
\selectlanguage{american}%

\affiliation{Department of Chemistry, University of Reading, Whiteknights, Reading
RG6 6AD, United Kingdom}
\selectlanguage{english}%

\author{Andrew I. Duff}
\selectlanguage{american}%

\affiliation{Scientific Computing Department, STFC Daresbury Laboratory, Hartree
Centre, Warrington, UK}
\selectlanguage{english}%

\date{\today}
\begin{abstract}
We have simulated the ultra-high-temperature ceramic zirconium carbide
(ZrC) in order to predict electron and phonon scattering properties,
including lifetimes and transport. Our predictions of heat and charge
conductivity, which extend to 3000 K, are relevant to extreme temperature
applications of ZrC. Mechanisms are identified on a first principles
basis that considerably enhance or suppress heat transport at high
temperature, including strain, anharmonic phonon renormalization and four-phonon scattering. The extent to which
boundary confinement and isotope scattering effects lower thermal
conductivity is predicted.
\end{abstract}

\maketitle

\section{Introduction\label{sec:Introduction}}

Zirconium carbide (ZrC) is a stiff material ($E\approx0.5$ GPa),\citep{Cheng2004,S2008}
with moderate volumetric mass density ($\rho=6.73$ g/cm\textsuperscript{3}),\citep{Lide}
metallic conductivity,\citep{Harrison2015} ultra-high melting-point
($T_{m}\approx3700$ K),\citep{FernandezGuillermet1995,Savvatimskiy2017}
and a low neutron scattering cross-section.\citep{Azevedo2011} Consequently,
ZrC is relevant to the development of advanced nuclear fuel clads
and aerospace materials. In each instance good thermal transport properties
are essential. Hypersonic flight for example requires sharp leading
edges capable of withstanding extreme-temperature ablative environments,\citep{Paul2014,Justin2011}
and in order to mitigate thermal shock, the material must be able
to effectively transport heat from the leading edge. Similarly in
the case of nuclear fuels, the clad must be able to deport thermal
energy effectively at high temperatures for safe reactor operation.

High-temperature conductivity measurements are non-trivial to make
so it is unsurprising that considerable scatter exists across the
reported thermal data.\citep{GROSSMAN1965,Harrison2015,Crocombette2013}
Our computational predictions are therefore expected to be of practical
use, as well as providing valuable theoretical insight into the basic
competing factors that determine high-temperature conductivity. Advances
in the computational treatment of phonons applied to thermal conductivity,\citep{Togo2015,Tadano2014,Li2014a}
provide us with a timely opportunity to present transport predictions
for the prototypical ultra-high-temperature ceramic ZrC.

In this paper we report DFT calculations on the electron and phonon
scattering of phonons in bulk ZrC. The results are used to predict
charge and heat transport, elucidating grain-size and isotopic effects,
as well as establishing the importance of accounting for strain, isobaric
heat enhancement, and strong high-temperature anharmonicity.

This paper is set out as follows. The theoretical background and technical
calculation details are listed in Sec. \ref{subsec:Theoretical-background}
and Sec. \ref{subsec:technicalDetails}. Results are presented in
two parts: phonon-phonon interactions and electron-phonon interactions
in Sec. \ref{subsec:Phonon-phonon-and-electron-phonon-interactions},
and heat and charge transport in Sec. \ref{subsec:Heat-and-charge-transport}.
Conclusions are presented in Sec. \ref{sec:Conclusions}, and additional
scattering information and methodological comparisons are given in
the Appendix.

\section{Methods\label{sec:Methods}}

\subsection{Theoretical background\label{subsec:Theoretical-background}}

The lattice contribution to conductivity is calculated within the
single-mode relaxation time approximation\citep{Chaput2011,Togo2015}

\begin{equation}
\kappa_{\text{ph}}(V_{0},T)=\frac{1}{N\,V_{0}}\sum_{\mu}C_{\mu}\mathbf{v}_{\mu}\otimes\mathbf{v}_{\mu}\tau_{\mu}^{\text{ph-ph}}\,,\label{eq: kappaPhonon}
\end{equation}
where $\tau_{\mu}^{\text{ph-ph}}$ is the relaxation time, $\mathbf{v}_{\mu}=\hbar\partial_{\mathbf{q}}\omega_{\mu}$
is the phonon velocity and $C_{\mu}=\hbar\omega_{\mu}\partial_{T}n_{\mu}^{\text{}}$
the heat capacity of band $\mu=n\mathbf{q}$, and $V_{0}$ is the
$T=0$ K volume of the cell at equilibrium strain.

The strain dependence of $\kappa_{\text{ph}}$ is usually neglected,
but in materials with large Grüneisen parameters or unusually high
melting points, thermal expansion can substantially affect
heat transport. In ZrC we account for the coupling to homogenous isotropic
principle axis strains by computing $\kappa_{\text{ph}}(V_{i},T)$
at a series of volumes spanning $[V_{0},\,V_{T_{m}}]$. The volume-dependent
conductivity $\kappa_{\text{ph}}(V,T)=\frac{1}{N\,V}\sum_{\mu}C_{\text{V}\mu}\left(V,T\right)\mathbf{v}_{\mu}\otimes\mathbf{v}_{\mu}\left(V\right)\tau_{\mu}^{\text{ph-ph}}\left(V,T\right)$
is calculated by a simple procedure that linearly mixes $\kappa_{\text{ph}}(V_{i},T)$
between pairs of volumes (temperatures) along the quasiharmonic thermal
expansion curve:

\begin{widetext}

\begin{equation}
\kappa_{\text{ph}}\left(V,T\right)=\sum_{i}\left\{ \kappa_{\text{ph}}\left(V_{i},T\right)\left(1-t_{i,\,i+1}\left(T\right)\right)+\kappa_{\text{ph}}\left(V_{i+1},T\right)t_{i,\,i+1}\left(T\right)\right\} \,\text{box}\left(T\right)\,.\label{eq: volumeDependentKappa}
\end{equation}

\end{widetext}$t_{i,\,i+1}(T)$ is a mixing parameter, $t_{i,\,i+1}(T)=\frac{T-T_{i}}{T_{i+1}-T_{i}}$,
that interpolates the temperature dependence of $\kappa_{\text{ph}}$
between the pairs of volumes (temperatures). The box function selects
the interpolation temperature range $[T_{i},\,T_{i+1}]$ as $\text{box}\left(T\right)=\Theta\left(T-T_{i}\right)-\Theta\left(T-T_{i+1}\right)$. 

As well as the effect of volume expansion on phonon thermal conductivity,
strong anharmonic effects become increasingly important at high temperature. We account for phonon frequency renormalization at finite temperature by explicitly considering anharmonicity up to fourth order using a recently developed real-space-based anharmonic phonon renormalization scheme.\citep{Xia2018a,Xia2018} The required high-order anharmonic interatomic force constants were constructed using compressive sensing lattice dynamics (CSLD).\citep{Zhou2014} Due to the computational expense, we only performed calculations at selected temperatures, for example, in the low (300 K), medium (1500 K), and high ($T_{m}$) temperature regimes. Moreover, we explicitly calculated the intrinsic phonon scattering rates from four-phonon processes, as recently formulated by Feng and Ruan,\citep{Feng2017} beyond the regularly used three-phonon scatterings. To estimate the impacts of anharmonic renormalization and four-phonon scattering on  conductivity at a range of temperatures,
the effects on $\kappa_{\text{ph}}(T)$ are interpolated in temperature
between the weakly and strongly anharmonic regimes, analogous to the
interpolation specified in Eqn. \ref{eq: volumeDependentKappa}.

The isobaric phonon conductivity $\kappa_{\text{ph}}(p)$ is calculated
by 
\begin{equation}
\kappa_{\text{ph}}(p)=\gamma\,\kappa(V)\,.\label{eq: isobaricKappaEffect}
\end{equation}
The enhancement factor, $\gamma=$ $C_{\text{P}}/C_{\text{V}}$, is determined
using $C_{\text{P}}$ computed at the quasi-harmonic level of theory.

Three-phonon relaxation lifetimes ($\tau_{\mu}^{\text{ph-ph}}=1/2\Gamma^{'',\text{ph-ph}}$)
are calculated based on the imaginary self-energies,
$\Gamma^{'',\text{ph-ph}}\equiv\text{Im}\,\Gamma^{\text{ph-ph}}$.\citep{Chaput2011,Togo2015}
$\Gamma^{'',\text{ph-ph}}$ is computed both by strain-dependent third-order
lattice dynamics,\citep{Chaput2011,Togo2015} and by CSLD to also account for high-temperature anharmonicity.\citep{Zhou2014}  Four-phonon scattering times are computed by iterative solution to the BTE,\citep{Xia2018a} at selected temperatures only ($300$ K, $1500$ K and $3800$ K). Three and four-phonon scattering times are combined using Matheisen's rule, and the total phonon relaxation lifetime including other terms is estimated as
\begin{equation}
\tau_{\mu}\text{=\ensuremath{\frac{1}{\sum_{i}1/\tau_{\mu}^{i}}\,,}}\label{eq: MatheisensRule}
\end{equation}
for $i\,\in\left\{ \text{ph-ph},\,\text{ph-iso},\,\text{ph-boundary}\right\} $.
Isotope mass defect scattering is treated perturbatively,\citep{Cardona2005a,Togo2015}
and boundary scattering as if providing a restriction on $\tau_{\mu}^{\text{ph-boundary}}=L/\mathbf{v}_{\mu}$
by domain size $L$.

The total thermal conductivity $\kappa_{\text{total}}$ is computed
as
\begin{equation}
\kappa_{\text{total}}=\kappa_{\text{ph}}+\kappa_{\text{el}}\,,\label{eq: totalThermalcond}
\end{equation}
with electron thermal conductivity $\kappa_{\text{el}}$ treated semi-classically\citep{Madsen2006,Pizzi2014}
\begin{equation}
\mathbf{\kappa}_{\text{el}}=\frac{1}{VT}\int d\varepsilon\,(\varepsilon-\mu)^{2}\,(-f')\sum_{\lambda}\overset{}{\mathbf{v}_{\lambda}}\otimes\overset{}{\mathbf{v}_{\lambda}}\tau_{\lambda}\delta(\varepsilon-\varepsilon_{\lambda})\,,\label{eq: electronKappa}
\end{equation}
and electrical conductivity is computed similarly\citep{Madsen2006,Pizzi2014}
\begin{equation}
\mathbf{\sigma}_{\text{el}}=\frac{1}{V}\int d\varepsilon\,(-f')\,\sum_{\lambda}\overset{}{\mathbf{v}_{\lambda}}\otimes\overset{}{\mathbf{v}_{\lambda}}\tau_{\lambda}\delta(\varepsilon-\varepsilon_{\lambda})\,,\label{eq: electronSigma}
\end{equation}
where $\lambda$ subsumes wavevector and band quantum numbers $\lambda\equiv\{n,\,\mathbf{k}\}$,
$f'=\partial_{\varepsilon}f(T,\varepsilon)$ is the electron occupancy
energy derivative, $\mathbf{v}_{\lambda}=\partial_{\mathbf{k}}\varepsilon_{n\mathbf{k}}$
is the band velocity tensor, and $\tau_{\lambda}$ is the electron relaxation time. 
The effect of thermal expansion on $\mathbf{\kappa}_{\text{el}}$ and $\mathbf{\sigma}_{\text{el}}$ is accounted for by computing each quantity for a series of volumes along the quasiharmonic thermal expansion curve. Conductivity tensors are determined, up to a factor of $\tau_{\lambda}$, from local density approximation (LDA) band
structures using Wannier functions, with methodological comparison
to Bloch functions and DFPT in Appendix.\citep{Mostofi2008,Mostofi2014,Pizzi2014,Kresse1996,Kresse1996a,Gajdos2006}
The relaxation time $\tau_{\lambda}$ is equated with the electron-phonon
scattering time $\tau_{\lambda}^{\text{el-ph}}$. $\tau_{\lambda}^{\text{el-ph}}$
is determined from $\tau_{\lambda}^{\text{el-ph}}=1/2\Sigma_{\lambda}^{'',\text{el-ph}}$,
with the imaginary part of the self-energy, $\Sigma_{\lambda}^{'',\text{el-ph}}\equiv\text{Im}\,\Sigma_{\lambda}^{\text{el-ph}}$,
found using the method of Poncé \emph{et al.}\citealp{Ponce2016,Giannozzi2009}.
This method is also used to compute $\Pi_{\mu}^{'',\,\text{el-ph}}$,
to determine the phonon relaxation time from electron-phonon scattering. 

\subsection{Technical details\label{subsec:technicalDetails}}

Three-phonon scattering rates are computed using the second-order
perturbation theory implemented by the \textsc{\footnotesize{PHONO3PY}}
code.\citep{Chaput2011,Togo2015} Small-displacement third-order force
constants are calculated at seven volumes that span the range
of thermal expansion. For third-order force constants
at each dilation, $144$ displacements are made on the $2\times2\times2$
($64$ atom) of ZrC supercell. Second-order force constant displacements
are made on a $4\times4\times4$ ($512$ atom) supercell at each
volume. \textbf{q}-points are sampled at a density equivalent to a
$31\times31\times31$ grid for the conventional eight atom unit cell.

Compressive sensing lattice dynamics (CSLD)\citep{Zhou2014} is used to account for strong anharmonic effects at high temperature, with force constant
tensors (FCTs) up to sixth order determined from snap-shots of uncorrelated quasi-random
configurations. Convergence is achieved for ZrC by twenty configurations of a $128$ atom supercell ($4\times4\times4$ of the primitive cell).

Force-constants are calculated using the \textsc{\small{}VASP} density
functional theory (DFT) code.\citep{Kresse1996,Kresse1996a} The PZ81
LDA\citep{Perdew1981a} functional provides a satisfactory description
of ZrC at low temperature, giving a zero-point-corrected $T=0$ K
value of $a_{\text{LDA}}=4.667$ Å, compared to an experimentally
reported value of $a_{\text{exp}}=4.694$ Å.\citep{Savvatimskiy2017}
The difference is similar in magnitude but opposite in sign to PBE,\citep{Perdew1996}
however we choose to work with the LDA due to reported superior description
of thermodynamics at high temperatures.\citep{Duff2015}

DFT calculations employ the projector-augmented wave (PAW) method,\citep{Kresse1999}
with $4$\emph{s} and $4$\emph{p}-Zr electron included as valence
states. Kinetic energy is cutoff above $700$ eV and \textbf{k}-points
are sampled at a density commensurate to a $12\times12\times12$ mesh
for the conventional cell. Methfessel-Paxton smearing is applied with
$0.2$ eV broadening.\citep{Methfessel1989} Cell total energies and
individual eigenvalues are converged to $10^{-8}$ eV, and force differences
to $10^{-6}$ eV/Å.

Electron-phonon lifetimes are calculated using the Electron-Phonon-Wannier
(\textsc{\small{}EPW}) code, interfaced with Quantum-Espresso\citealp{Ponce2016}
(\textsc{\small{}QE}).\citealp{Giannozzi2009} \textbf{k}-space interpolation
uses maximally-localized Wannier functions generated using\textsc{\small{}
WANNIER90}.\citealp{Mostofi2008,Mostofi2014,Marzari1997,Marzari2012a,Souza2002} 

\textsc{\small{}QE} calculations use an LDA exchange-correlation functional,\citealp{Perdew1981a,Ceperley1980a}
with a projected augmented wave (PAW) pseudopotential for Zr with
4\emph{s}\textsuperscript{2}4\emph{p}\textsuperscript{6}4\emph{d}\textsuperscript{2}5\emph{s}\textsuperscript{2}
electrons considered as valence electrons, and a norm-conserving pseudopotential
for C used with 2\emph{s}\textsuperscript{2}2\emph{p}\textsuperscript{2}
electrons treated as valence electrons. Convergence of $0.5$ mRy
is obtained with a \textbf{$12\times12\times12$ }$\Gamma$-centered
\textbf{k}-point mesh and a kinetic energy cutoff of $200$ Ry. Ionic
minimization is performed until energy differences are less than $10^{-6}$
Ha and force differences less than $10^{-5}$ Ha/Bohr. Electronic
convergence is at least $10^{-10}$ Ry.

Electron-phonon interaction strengths are found using dynamical matrices
from DFPT.\citealp{Baroni1987,Baroni2001a,Gonze1997} An irreducible
$\Gamma$-centered $6\times6\times6$ \textbf{q}-point mesh is used
with convergence criteria of at least $10^{-14}$ Ry. Non-self-consistent
calculations are performed on a coarse $\Gamma$-centered \textbf{$6\times6\times6$
k}-point mesh using the same energy criteria as the energy minimization.
Both the coarse \textbf{q}-point and \textbf{k}-point meshes were
tested for convergence. These results are then used for Wannier interpolation. 

The Wannier functions are projected onto carbon \emph{sp}\textsuperscript{3}-orbitals
and three Zr \emph{d}-orbitals.  Four bands below the Fermi level representing the Zr
\emph{s}-band and the three Zr \emph{p}-bands are not included in
the calculation as well as the highest conduction band. In each calculation
the disentangled method is used \citep{Souza2002} and the disentangled
energy window is set to between $12$ eV below the Fermi level to
include the C \emph{s}-orbital and $2$ eV above the Fermi level.
These settings provide the best  spreads (between $1.5\,\lyxmathsym{\textendash}\,1.7$
Å\textsuperscript{2} per Wannier function). As Wannier interpolation
on homogeneous \textbf{k} and \textbf{q}-grids of $45\times45\times45$
were unable to achieve convergence, randomly generated grids were used
as suggested by Poncé \emph{et al}..\citealp{Ponce2016} A grid of
$50,000$ randomly generated \textbf{k}-points and $150,000$ randomly
generated \textbf{q}-points with a broadening of $20$ meV is sufficient
for convergence. 

The conductivity tensor in Eq. \ref{eq: electronSigma} is determined
from \textsc{\small{}VASP}-calculated\citep{Kresse1996,Kresse1996a}
LDA\citep{Perdew1981a} band structures. The technical parameters
are identical to the \textsc{\small{}VASP} phonon calculation details
described previously except for the reciprocal-space sampling density.
The electronic band structure calculations employed a dense \textbf{k}-point
mesh sampling of $39\times39\times39$, from which the conductivity
tensor is determined by i) linear response routines in \textsc{\small{}VASP},\citep{Gajdos2006}
ii) reciprocal-space band velocities \emph{via} \textsc{\small{}BoltzTraP},\citep{Madsen2006}
and iii) by real-space band velocities \emph{via} \textsc{\small{}BoltzWann}.\citep{Mostofi2008,Mostofi2014,Pizzi2014}
Electrical transport tensors account for the thermal expansion
of the lattice although the effect is marginal.

\section{Results\label{sec:Results}}

\begin{figure*}[t]
\begin{raggedright}
\includegraphics[scale=0.47]{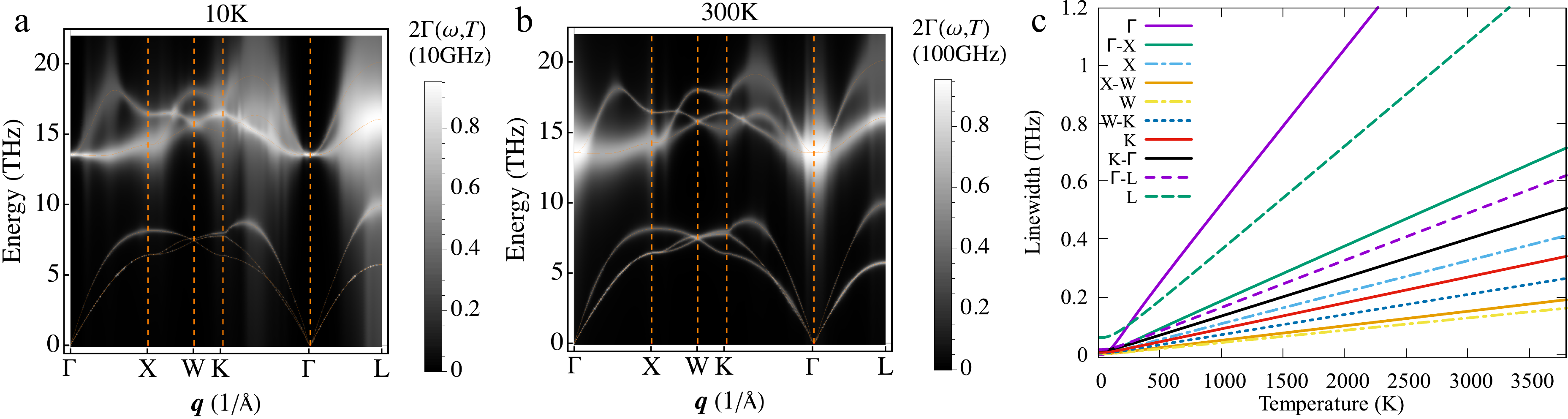}
\par\end{raggedright}
\caption{a-b) Interaction broadened harmonic phonon dispersion at 10 K and
300 K. FWHM $\Gamma_{\lambda}^{'',\text{ph-ph}}(\omega,\mathbf{q},T)$
line broadening is Lorentzian and specified by the contour brightness
scale on the right axis which ranges from $0$-$10$ GHz for $10\,\text{K}$
to $0$-$100$ GHz for $300\,\text{K}$. c) Linewidth \emph{vs} temperature
at selected points (labelled) in the Brillouin zone. \label{fig: ph-ph-LineWidths}}
\end{figure*}

\subsection{Phonon-phonon and electron-phonon scattering\label{subsec:Phonon-phonon-and-electron-phonon-interactions}}

The acoustic-type phonon bands in ZrC, which project more than $98$
\% onto the motion of Zr atoms, are weakly interacting compared to
the carbon-projecting optic-type phonon dispersion bands. This is
shown for ZrC at 10 K and at 300 K by the phonon-phonon interaction
broadened dispersion in Fig. \ref{fig: ph-ph-LineWidths}a-b. Optic-type
modes at $\mathbf{q}=0$ are the most strongly interacting for $T>300$
K. For $T<300$ K, other lower symmetry wavevectors, for instance
between $K\text{-}\Gamma$ and $L\text{-}\Gamma$ in the Brillouin
zone, become more actively scattering than the zone-center $\Gamma$-point
optic modes in Fig. \ref{fig: ph-ph-LineWidths}a-b. 

The linear temperature dependence of the ZrC linewidths is shown for
a range of wavevectors in Fig. \ref{fig: ph-ph-LineWidths}c, with
average values comparable to those reported for other materials.\citep{Togo2015}
For example the average three-phonon linewidth is $0.05$ THz at $300$
K, which is lower than the CuCl 300 K linewidth ($1.3$ THz) and slightly
larger than GaAs ($0.03$ THz). In ZrC we emphasize the limited insight
from quoting $n\mathbf{q}$ averaged values, as the linewidths are
quite strongly anisotropic (as pictured, Fig. \ref{fig: ph-ph-LineWidths}b). On the validity of the perturbative quasi-particle formalism within
which lifetimes are determined, acoustic modes are weakly interacting
to $T_{m}$ but caution is necessary on the interpretation of optic
modes at high-symmetry points for which linewidths can be of the order
1 THz for temperatures exceeding $0.6\,T_{\text{m}}$ .

\begin{figure*}[tp]
\includegraphics[scale=0.9]{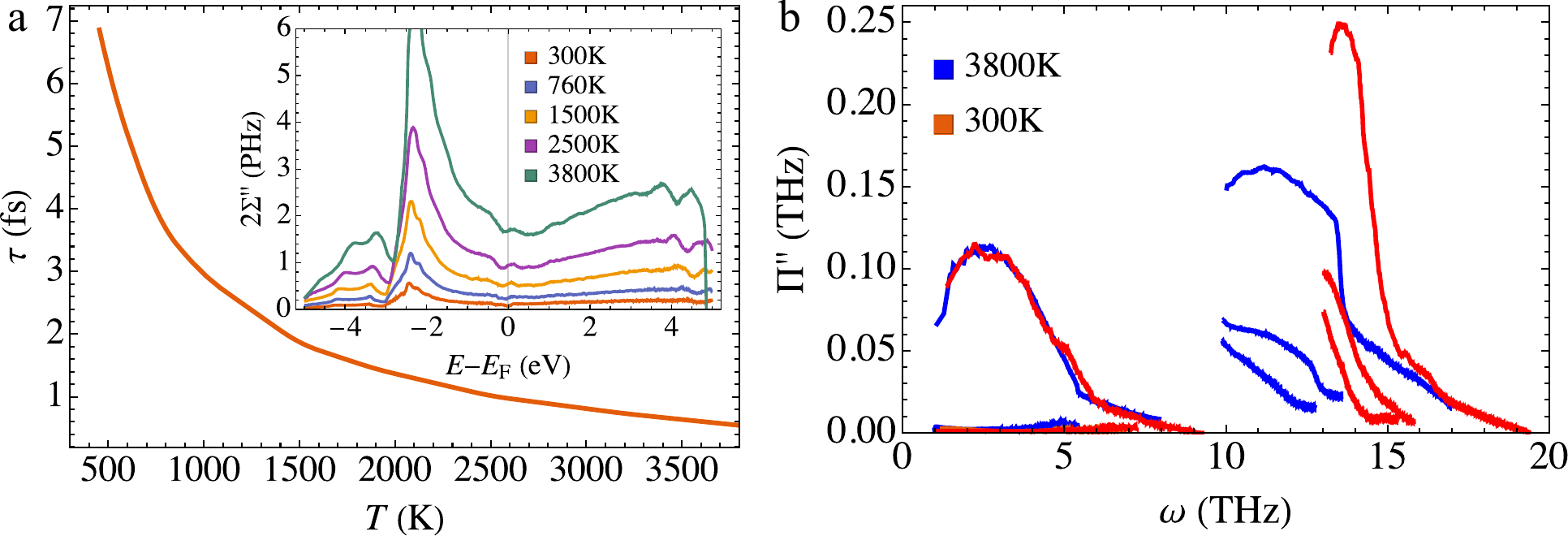} 
\caption{a) Electron relaxation time and associated imaginary self-energy $\Sigma_{\lambda}^{'',\text{el-ph}}(\varepsilon,T)$
\emph{inset}. b) Imaginary phonon self-energy $\Pi^{'',\text{el-ph}}(\omega,T)$,
for each of the three acoustic and optic bands.\label{fig: electronPhononSelfEnergies}}
\end{figure*}

ZrC is unusual in exhibiting ceramic and metallic bonding characteristics.
To describe the electrical conductivity of  ZrC we account for electron-phonon
scattering. The average time for the scattering of electrons by phonons, calculated by
\begin{equation}
\tau^{\text{el-ph}}(T)=\int d\varepsilon\,\frac{1}{N_{n\mathbf{k}}}\sum_{n,\mathbf{k}}\tau_{n\mathbf{k}}^{\text{el-ph}}(T)\delta(\varepsilon-\varepsilon_{n,\mathbf{k}})f'(\varepsilon,T),\label{eq: electron-Phonon-relaxationTime}
\end{equation}
is shown in Fig. \ref{fig: electronPhononSelfEnergies}a. The average
relaxation time decreases with temperature, for example from $\tau^{\text{el-ph}}(500\,\text{K})=6.3\,\text{fs}$
to $\tau^{\text{el-ph}}(2500\,\text{K})=1\,\text{fs}$. As ZrC is
semi-metallic, $\tau^{\text{el-ph}}(T)$ is enhanced with temperature
as $f'=\partial_{\varepsilon}f(T,\varepsilon)$ samples states that
increase in concentration away from the Fermi energy. This is demonstrated
in Fig. \ref{fig: electronPhononSelfEnergies}a-inset\emph{ }in the
electron self-energy $\Sigma^{'',\text{el-ph}}$ which is a local
minimum about the Fermi energy. Recent work has shown intrinsic defects
such as Frenkel defects, which are predicted to spontaneously generate at high temperature in ZrC, can as much double the density of states at the Fermi
energy.\citep{Mellan2018} This is likely to modulate $\Sigma^{'',\text{el-ph}}$ and electron transport at high temperature, though
explicit characterisation with first principles calculations is beyond
the scope of this work. 

In experiments the width of a phonon line may be measurable, while
the origin of the broadening remains obscured, so it is interesting
to compare phonon linewidths from electron-phonon interactions $2\Pi^{'',\text{el-ph}}$
and from three-phonon interactions $2\Gamma^{'',\text{ph-ph}}$.
Values for $2\Gamma^{'',\text{ph-ph}}$ and $2\Pi^{'',\text{el-ph}}$
are listed for selected temperatures in Table \ref{tab:Phonon-linewidths}.
At $300$ K, $2\Gamma_{\mathbf{q=0},\text{optic}}^{'',\text{ph-ph}}$
peak values are smaller (ca. $\times3$) than $2\Pi_{\mathbf{q=0},\text{optic}}^{'',\text{el-ph}}$,
but for higher temperatures phonon-phonon interaction is many times
greater than electron-phonon (ca. $\times8$). As a crystal with partially
occupied states at the Fermi energy, it is a point of interest that the room
temperature total phonon linewidth has comparable contributions from
anharmonic and electron-phonon interactions. This observation for the linewidth in conducting
crystalline systems is however not exceptional, and has recently
been reported at room temperature in a number of systems, including
graphite,\citep{Paulatto2015,Bonini2007} noble metals with small
DOS(\emph{E}\textsubscript{F}) such as Cu, Ag and Au,\citep{Tang2011,Bauer1998,Lin2008}
as well as more exotic systems such as the superconductor palladium
hydride.\citep{Paulatto2015} 

\begin{table}
\caption{Phonon linewidths from electron-phonon and phonon-phonon interactions.
\label{tab:Phonon-linewidths}}

\begin{centering}
\begin{tabular}{ccc}
\hline 
\noalign{\vskip\doublerulesep}
\emph{T} & $2\Pi_{\mathbf{q}=0,\text{optic}}^{'',\text{el-ph}}$ & $2\Gamma_{\mathbf{q}=0,\text{optic}}^{'',\text{ph-ph}}$\tabularnewline[\doublerulesep]
\hline 
\noalign{\vskip\doublerulesep}
\hline 
\noalign{\vskip\doublerulesep}
300 K & 0.50  THz (17  cm\textsuperscript{-1}) & 0.17 THz (5.7 cm\textsuperscript{-1})\tabularnewline
\noalign{\vskip\doublerulesep}
\noalign{\vskip\doublerulesep}
3000 K & 0.32  THz (11  cm\textsuperscript{-1}) & 2.4 THz (80 cm\textsuperscript{-1})\tabularnewline[\doublerulesep]
\hline 
\noalign{\vskip\doublerulesep}
\end{tabular}
\par\end{centering}
\centering{}
\end{table}

\subsection{Heat and charge transport\label{subsec:Heat-and-charge-transport}}

\begin{figure*}
\includegraphics[scale=0.74]{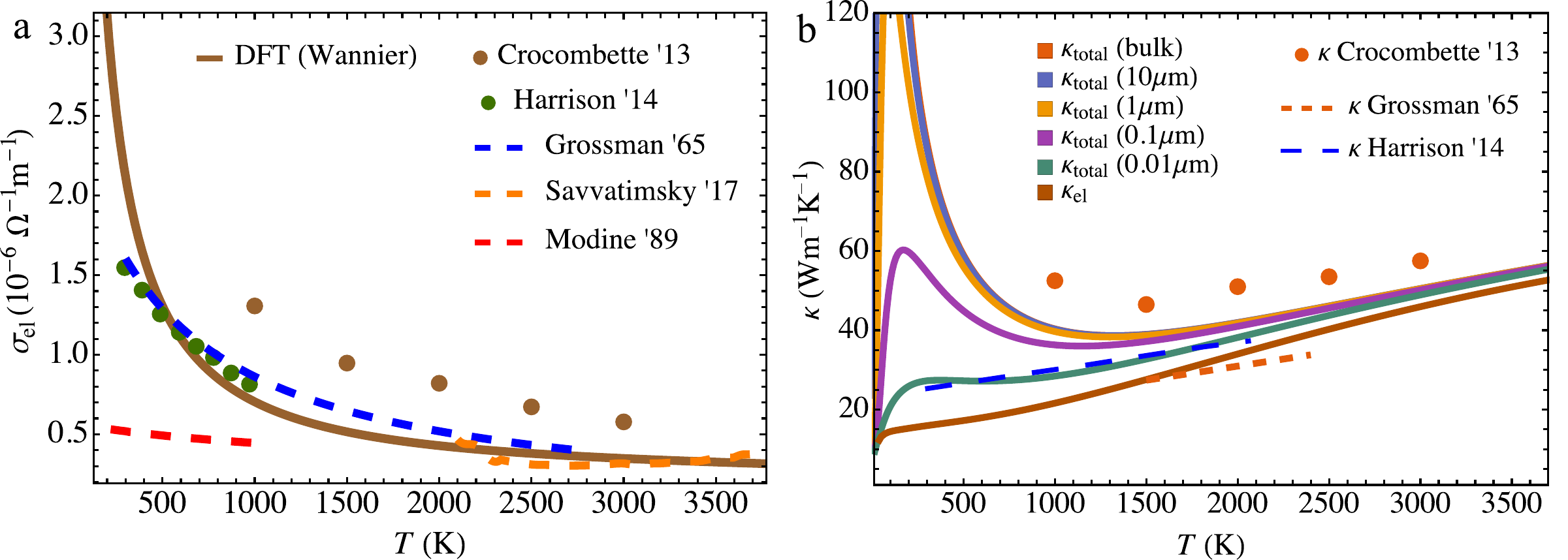}
\caption{a) Electrical conductivity $\sigma_{\text{el}}$ predictions (solid
line). b) Thermal conductivity predictions for $\kappa_{\text{total}}=\kappa_{\text{ph}}+\kappa_{\text{el}}$
and $\kappa_{\text{el}}$. Dashed and dotted lines denote
the experimental measurements\citep{Harrison2015,GROSSMAN1965,Savvatimskiy2017,Modine1989}
and computed results of Crocombette.\citep{Crocombette2013}\label{fig: kappaTotal}}
\end{figure*}

Unlike most ceramics ZrC is a relatively good conductor of electricity.\citep{Harrison2015,Modine1989,GROSSMAN1965,Savvatimskiy2017}
We have calculated the electrical conductivity ($\sigma_{\text{el}}$) of ZrC
within a first principles Boltzmann transport approach, using the Wannier function DFT method
to determine the electron-phonon relaxation time and band velocities. This method
is compared to other approaches in the Appendix. The predicted values
of $\sigma_{\text{el}}$ range from $\sigma_{\text{el}}(500\,\text{K})=1.4\times10^{-6}\,\Omega^{-1}\text{m}^{-1}$
to $\sigma_{\text{el}}(2500\,\text{K})=3.8\times10^{-7}\,\Omega^{-1}\text{m}^{-1}$,
with $\sigma_{\text{el}}(T)$ shown in Fig. \ref{fig: kappaTotal}a\emph{, }alongside experimentally measured values from multiple sources.\citep{Harrison2015,GROSSMAN1965,Modine1989,Savvatimskiy2017}
The electron thermal conductivity $\kappa_{\text{el}}(T)$
is shown in Fig. \ref{fig: kappaTotal}b. Starting at zero temperature $\kappa_{\text{el}}(T)$
exhibits a sharp increase initially, then increases almost linearly, rising from $\kappa_{\text{el}}(500\,\text{K})=17\,\text{W}\text{m}^{-1}\text{K}^{-1}$
to $\kappa_{\text{el}}(2500\,\text{K})=40\,\text{W}\text{m}^{-1}\text{K}^{-1}$.

In each instance the $\sigma_{\text{el}}$ and $\kappa_{\text{el}}$ values in Fig. \ref{fig: kappaTotal}a-b computed from first principles give reasonable agreement with experiment. This even appears to be true as the system nears the Mott-Ioffe-Regel limit where resistivity usually saturates.\citep{Werman2016, Modine1989, Auerbach1984, Werman2017, Sundqvist2014} The physical origin of additional transport channels leading to saturation can be somewhat debatable but is often well-described in an empirical sense by a parallel shunt model. It is therefore likely our high-temperature conductivity predictions are a lower limit for the conductivity of the defect-free crystal. The reasonable agreement of our transport predictions with experimentally reported ones at high temperatures is partially explained by electron quasi-momentum quantum numbers that remain moderately good up to quite high temperatures -- for example, at 3000 K the electron mean-free path is still ca. $\times 3\,a$ lattice parameters. Above this temperature the electron transport plots in Fig. \ref{fig: kappaTotal}a-b are extended for reference only and with underlined caution due to the inadequacy of the Boltzmann electron transport picture close to $T_m$ when scattering lengths and lattice parameters become close.

The total phonon and electron thermal conductivity, $\kappa_{\text{total}}=\kappa_{\text{ph}}+\kappa_{\text{el}}$,
is shown \emph{versus }temperature in Fig. \ref{fig: kappaTotal}b\emph{. }Whether $\kappa_{\text{ph}}$ or $\kappa_{\text{el}}$ is
the larger contribution to $\kappa_{\text{total}}$ depends on temperature
and other factors such as geometric constraints. For example, grain
boundaries introduce a restriction on maximum phonon path length for
weakly scattering Zr modes that considerably changes $\kappa_{\text{ph}}$.
Large domains ($L\geqq10\,\mu\text{m}$) permit the transport of heat
by long-wavelength high-velocity modes in ZrC, resulting in high values
such as $\kappa_{\text{total}}(300\,\text{K})=87\,\text{W}\text{m}^{-1}\text{K}^{-1}$.
Grains of moderate sizes
suppress phonon transport at low temperatures, with our model predicting
$\kappa_{\text{total}}(300\,\text{K})=55\,\text{W}\text{m}^{-1}\text{K}^{-1}$ at $L=0.1\,\mu\text{m}$.
Small domain sizes such as $L=0.01\,\mu\text{m}$ severely
limit transport processes from
weakly scattering high-velocity modes, further lowering $\kappa_{\text{total}}(300\,\text{K})$ to $27\,\text{W}\text{m}^{-1}\text{K}^{-1}$.
The size dependence of $\kappa_{\text{total}}$ suggests grain control
by sintering or synthesis temperature is important to design the transport behavior of ZrC. More details on grain boundary scattering
and the extent of mean-free path saturation expected at high temperature are given in the Appendix. 

In Fig. \ref{fig: kappaEffects} we show a range of temperature-dependent
mechanisms that can enhance or suppress phonon thermal conductivity.
Note, each effect that is analyzed has been included the prior calculation
of $\kappa_{\text{total}}$ that was presented Fig. \ref{fig: kappaTotal}b. 

Dilation of ZrC by tensile principal axis strains is found to lower
$\kappa_{\text{ph}}$ considerably. For instance thermal expansion
decreases phonon conductivity by up to 65\% relative to $\kappa_{\text{ph}}$  at the 0-K equilibrium volume, as shown by the enhancement factor $\kappa_{\text{qha}}^{\text{ph}}[V(T)]/\kappa_{\text{}}^{\text{ph}}(V_{0})$
\emph{versus} temperature in Fig. \ref{fig: kappaEffects}a. One
way to rationalize the large change is by considering how the stiffness
(specifically, isothermal bulk modulus $K_{\text{T}}$) and volumetric
mass density ($\rho$) change with temperature relative to each other.
Provided $\left|\frac{\partial K_{\text{T}}}{\partial T}\right|/\left|\frac{\partial\rho}{\partial T}\right|<1$, $\kappa_{\text{ph}}$ decreases with thermal expansion. Indeed this
is the case and a decrease of $39$ \% is observed from $T=10\,\text{K}$
to $T_m$. As $K_{\text{T}}/\rho$ is equal to the long
wavelength-limit band velocity $v_{s}^{2}$, the decrease in $\sqrt{K_{\text{T}}/\rho}$
is equivalent to decrease in acoustic band velocity from $52.1$ THzÅ
($5210$ m/s) to $44.1$ THzÅ ($4410$ m/s). This drop accounts for
a considerable part of the decrease in $\kappa_{\text{ph}}$ with
temperature in Fig. \ref{fig: kappaEffects}a.
 
Quantum zero-point motion modifies the equilibrium configuration
through tensile strain. In ZrC this softens modes and decreases $\kappa_{\text{ph}}$
by a temperature-independent factor of ca. 5\%, which is shown in
Fig. \ref{fig: kappaEffects}b. Conversely the isobaric heat capacity
enhancement factor (see Eqn. \ref{eq: isobaricKappaEffect}) is negligible
at low temperature, but increases $\kappa_{\text{ph}}$ by as much
as 35 \% at high temperature, as shown in Fig. \ref{fig: kappaEffects}f. 

At low temperature mass scattering at the natural isotopic abundance
in ZrC is important. For instance $\kappa_{\text{ph}}$ in Fig. \ref{fig: kappaEffects}c is more than 50 \% lower than an artificially prepared isotopically
pure crystal at low temperature, but effect quickly falls off with
increasing temperature. Grain size effects are also substantial at
low temperature, as illustrated in the prior discussion mean-free path constraints for Fig. \ref{fig: kappaTotal}b. For length scale effects from the opposite limit, minimum phonon conductivity saturation\citep{Sun2010, Slack1979} is expected to be comparatively smaller even at high temperatures. This is based on the accumulated phonon thermal conductivity as a function of mean-free path, which is given in Fig. \ref{fig: mfpEffects}, in the Appendix.

For high temperatures ($>0.5\,T_{\text{m}}$) perturbative third-order
lattice dynamics becomes insufficient as the system explores atomic
displacements far from the equilibrium configuration. Strong anharmonicity
renormalizes dispersion bands to higher frequencies, opposite to the
$\frac{\partial\text{ln}\left|\omega\right|}{\partial\text{ln}V}<0$
typical volume softening of frequencies with positive thermal
expansion. The high-temperature anharmonic renormalization considerably
enhances $\kappa_{\text{ph}}$ as shown in Fig. \ref{fig: kappaEffects}d. The effect as a fraction of the high-temperature conductivity can be large, for example increasing $\kappa_{\text{ph}}$ by ca. 100 \% for temperatures in excess of 2000 K. 

It has recently been noted that four-phonon scattering plays a critical role in determining the lattice thermal conductivities in both weakly and strongly anharmonic systems, e.g., BAs\citep{Feng2017} and PbTe\citep{Xia2018a} respectively. In ZrC we observe that four-phonon scattering processes become very prominent at high temperature for $\kappa_{\text{ph}}$. As with PbTe,\citep{Xia2018a} renormalization enhances $\kappa_{\text{ph}}$ and four-phonon processes suppress $\kappa_{\text{ph}}$.  The degree of cancellation between these anharmonic effects can be observed by considering Fig. \ref{fig: kappaEffects}d-e, and is illustrated by noting the renormalization enhancement of $\kappa_{\text{ph}}$ at 300 K (3800 K) is +4.2 W/mK (+6.2W/mK), while four-phonon scattering lowers $\kappa_{\text{ph}}$ by -4.1 W/mK (-5.8 W/mK). Additional details on four-phonon scattering and anharmonic phonon renormalization are given in the Appendix. 

\begin{figure}
\includegraphics[scale=0.63]{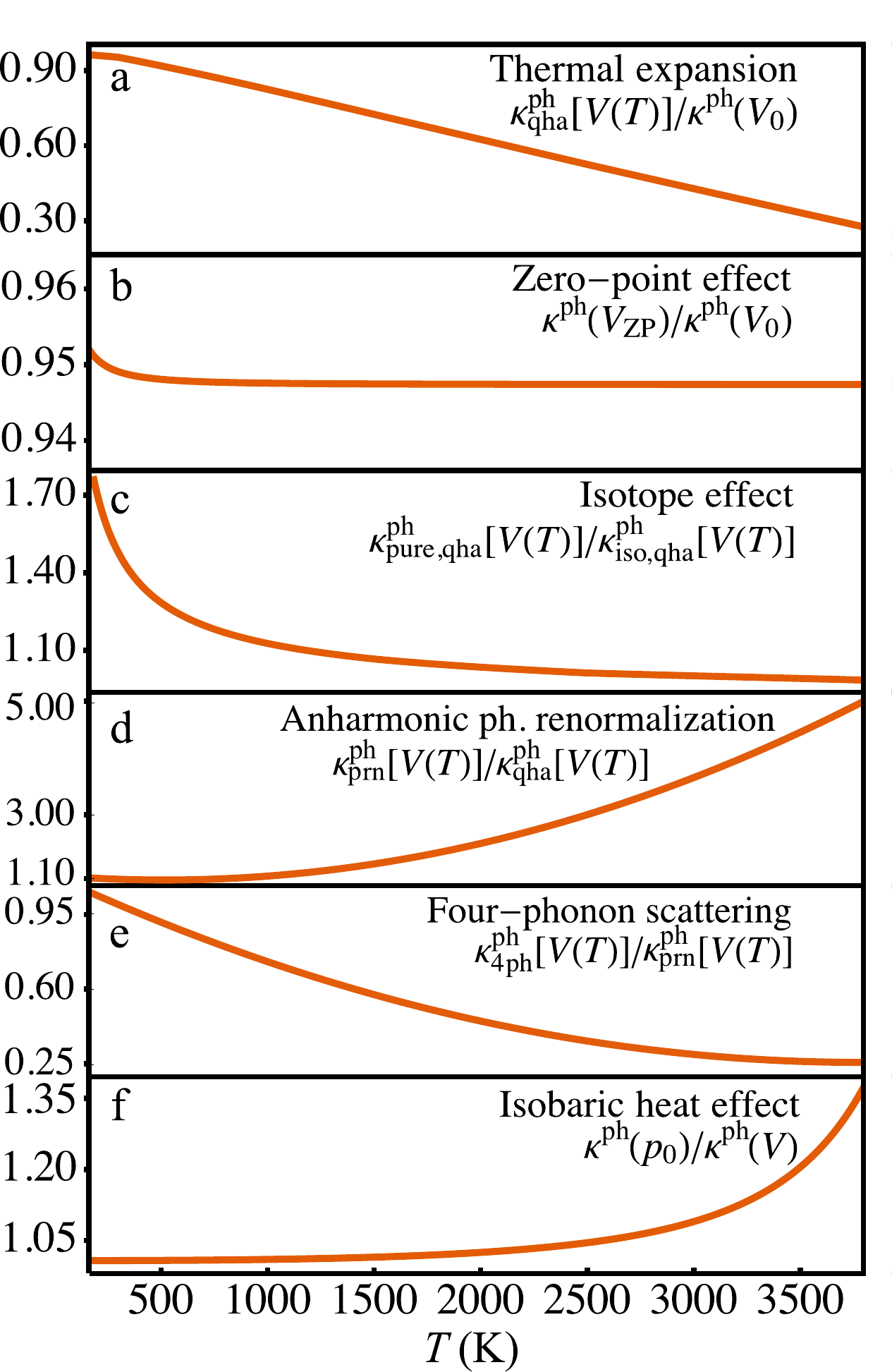} 
\caption{Phonon thermal conductivity enhancement and suppression mechanisms
in ZrC. a) Thermal expansion lowers conductivity by $\kappa_{\text{qha}}^{\text{ph}}[V(T)]/\kappa_{\text{}}^{\text{ph}}(V_{0})$.
b) Zero-point motion lowers conductivity by $\kappa_{\text{}}^{\text{ph}}(V_{\text{ZP}})/\kappa_{\text{}}^{\text{ph}}(V_{0})$.
c) Isotope purity enhances conductivity by factor of $\kappa_{\text{pure,qha}}^{\text{ph}}[V(T)]/\kappa_{\text{iso,qha}}^{\text{ph}}[V(T)]$
compared mass scattering at the natural isotopic abundance. d) Anharmonic phonon renormalization enhances conductivity by $\kappa_{\text{prn}}^{\text{ph}}(V)/\kappa_{\text{qha}}^{\text{ph}}(V)$ relative to a crystal with quasiharmonic frequency dependence.
e) Additionally including four-phonon scattering suppresses conductivity by $\kappa_{\text{4ph}}^{\text{ph}}(V)/\kappa_{\text{prn}}^{\text{ph}}(V)$.
f) Isobaric heat effect enhances conductivity $\kappa^{\text{ph}}(p_{0})/\kappa_{\text{}}^{\text{ph}}(V)$.
\label{fig: kappaEffects}}
\end{figure}

\section{Conclusions\label{sec:Conclusions}}

We have reported first principles calculations on the scattering and
transport properties of electrons and phonons in the ultra-high-temperature
ceramic ZrC. The nature of the phonon linewidth in ZrC has been examined
in terms of the energy dependence, anisotropy across the Brillouin
zone and temperature dependence of phonon-phonon and electron-phonon
interactions. In each instance phonons primarily scatter \emph{via}
optic modes at the $\Gamma$ point. The total phonon linewidth is
predominantly phonon-phonon in character rather than electron-phonon
for all but the lowest temperatures.

The electrical and thermal conductivities $\sigma_{\text{el}}$ and
$\kappa_{\text{el}}$ have been predicted at ambient pressure as a
function of temperature, along with total thermal conductivity $\kappa_{\text{total}}=\kappa_{\text{el}}+\kappa_{\text{ph}}$.
Thermal expansion of crystal volume notably suppresses thermal conductivity,
decreasing the phonon contribution by more than 50\% at $0.75\,T_{\text{m}}$.
Suppression of thermal conductivity by strain sources, such as thermal
expansion and grain boundaries, should be considered when engineering
heat dissipation of an ultra-high temperature ceramic for extreme
environment applications. At low temperature $\kappa_{\text{ph}}$
is considerably lowered by isotope mass defect scattering, and by
features that enforce geometric constraints such as grain boundaries
that prevent the crystal supporting long-lived phonon modes. Isobaric heat capacity, and anharmonic phonon renormalization,
provide substantial enhancements in $\kappa_{\text{ph}}$ at high temperature (ca. $0.75\,T_{\text{m}}$). Four-phonon scattering strongly suppresses thermal conductivity at high temperature, almost cancelling the anharmonic frequency renormalization effect.

The examination of point and extended defects and sub-stoichiometry
on transport is beyond the scope of this work, but would be a valuable
future extension to this work. As would the examination of saturation effects and non-quasiparticle transport for $T>0.75\,T_m$.

\section{Acknowledgements\label{sec:Acknowledgements}}

T.A.M. acknowledges the financial support of EPSRC Programme Grant
Material Systems for Extreme Environments (XMat) (Grant No. EP/K008749/2),
EPSRC Programme Grant Carbides for Future Fission Environments (CAFFE)
(Grant No. EP/M018563/1), and H2020 project Il Trovatore (Grant No.
740415). T.A.M. acknowledges computational resources from the UK Materials
and Molecular Modelling Hub (Grant No. EP/P020194/1) and embedded
CSE 33 of the ARCHER UK National Supercomputing Service (\href{http://www.archer.ac.uk}{http://www.archer.ac.uk}).
\foreignlanguage{american}{A.I.D. acknowledges support from the STFC
Hartree Centre\textquoteright s Innovation: Return on Research programme,
funded by the UK Department for Business, Energy \& Industrial Strategy.}
\break

\section{References\label{sec:References}}

\selectlanguage{american}%
\bibliographystyle{apsrev4-1}

\bibliography{ZrC_transport_manuscript}

\appendix

\section*{Appendix}

\section*{Methodological comparisons\label{sec:Methodological-comparisons}}

\begin{figure}
$\,$

\includegraphics[scale=0.9]{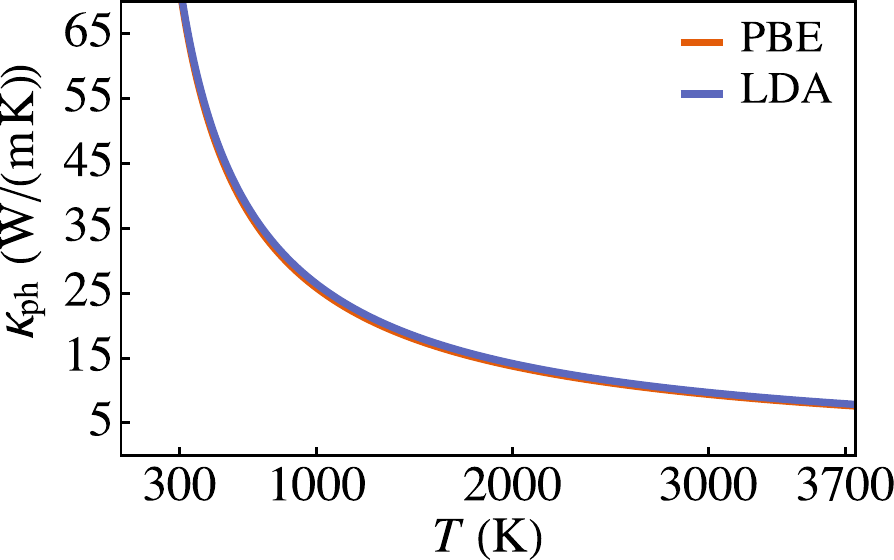}
\caption{Comparison of the ZrC phonon thermal conductivity computed with the
PBE and LDA exchange-correlation functionals at $a=4.667$ Å. \label{fig: kappaLDAPBE}}
\end{figure}

\selectlanguage{english}%
The ZrC phonon conductivity calculated using the LDA and PBE exchange-correlation
functionals are shown in Fig. \ref{fig: kappaLDAPBE}. At fixed equal
volumes, the difference in phonon conductivity is negligible. 

\begin{figure}
\begin{raggedright}
\includegraphics[scale=0.85]{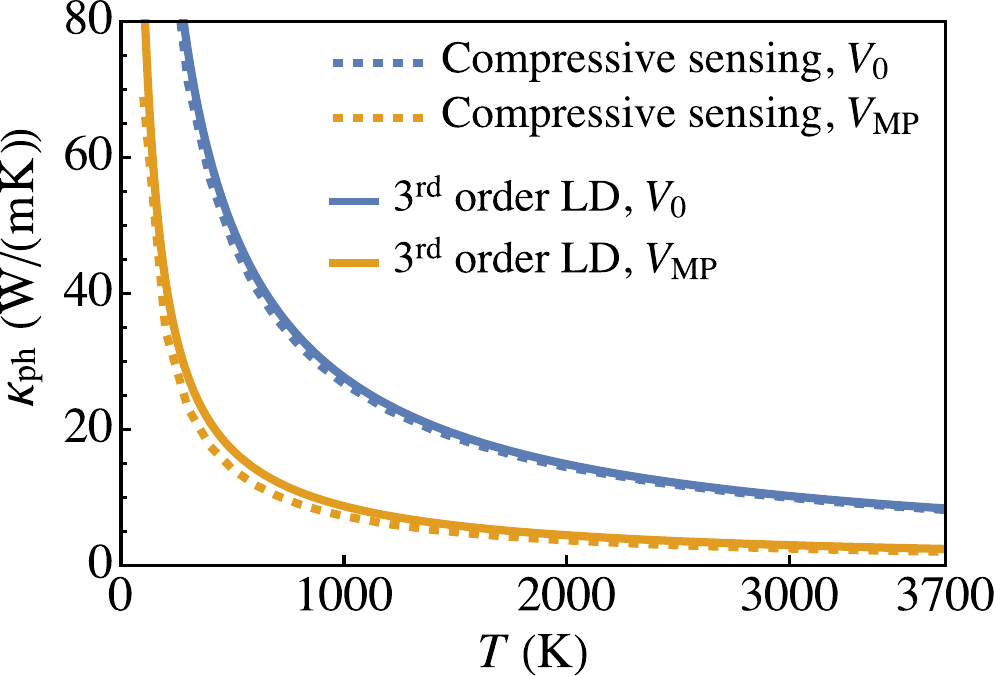}
\par\end{raggedright}
\caption{Comparison of conductivity from compressive sensing\citep{Zhou2014}
and third-order lattice dynamics (Phono3py\citep{Togo2015}), at zero-temperature
equilibrium and high-temperature volumes.\label{fig: harmonicCSPhono3py}}
\end{figure}

In Fig. \ref{fig: harmonicCSPhono3py} the phonon thermal conductivities
are compared from the compressive sensing\citep{Zhou2014} and third-order\citep{Togo2015}
lattice dynamics approaches. Compressive sensing predicts a similar
but marginally smaller thermal conductivity.

\begin{figure}
\includegraphics[scale=0.8]{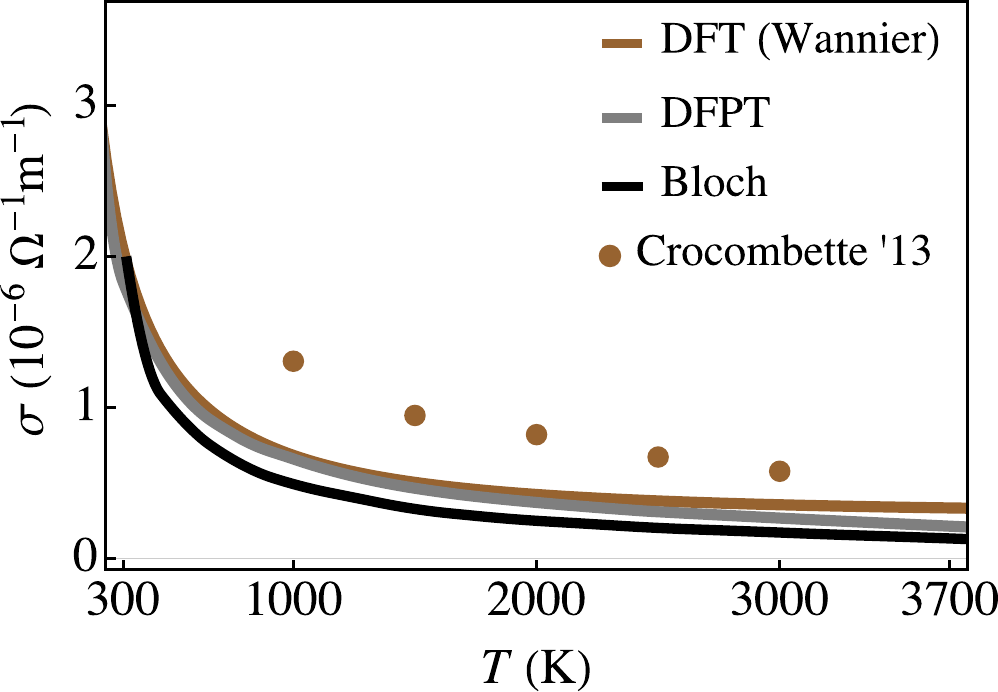}
\caption{Comparison of computational approaches to determine electrical conductivity.
1) LDA + Wannier functions, 2) LDA + DFPT linear response, 3) LDA
+ Bloch functions, and 4) PBE + snapshots from classical potential
MD\citep{Crocombette2013}. \label{fig: comparison-sigma-el}}
\end{figure}

The electrical conductivity from Wannier function, Bloch function
and DFPT LDA DFT calculations is shown in Fig. \ref{fig: comparison-sigma-el}.
The conductivity with Wannier functions is expected to provide the
most accurate predictions.

\section*{Phonon scattering geometric constraints\label{sec:Phonon-scattering}}
\begin{figure}
\includegraphics[scale=0.7]{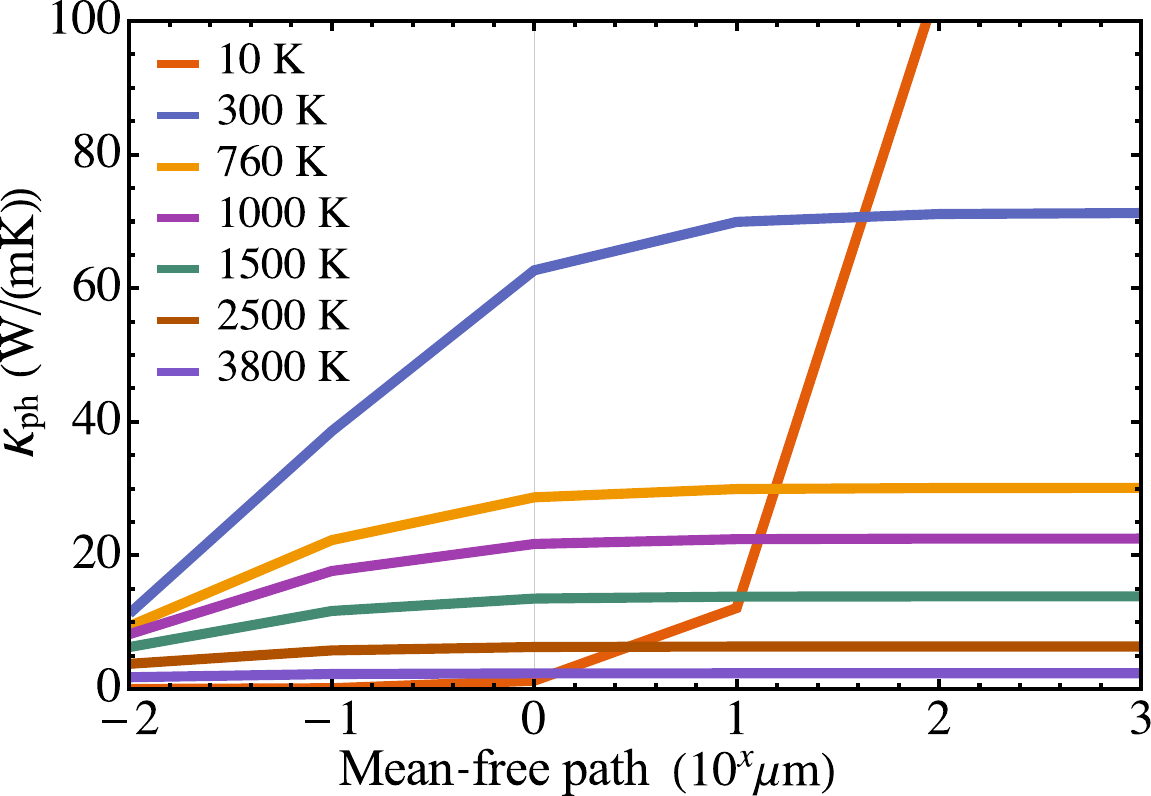}
\caption{Grain-size effects estimated by geometric restrictions on phonon conductivity
at a series of temperatures in perfect ZrC. Phonon cutoff lengths
span the interval $[10^{-2},\,10^{3}]\,\mu\text{m}$. \label{fig: boundaryEffects} }
\end{figure}

The effect of size constraints on phonon thermal conductivity is shown
in Fig. \ref{fig: boundaryEffects}. The geometry restriction effect
is strongly temperature dependent, lowering the ZrC phonon conductivity
most acutely at low temperature.

\begin{figure}
\includegraphics[scale=0.6]{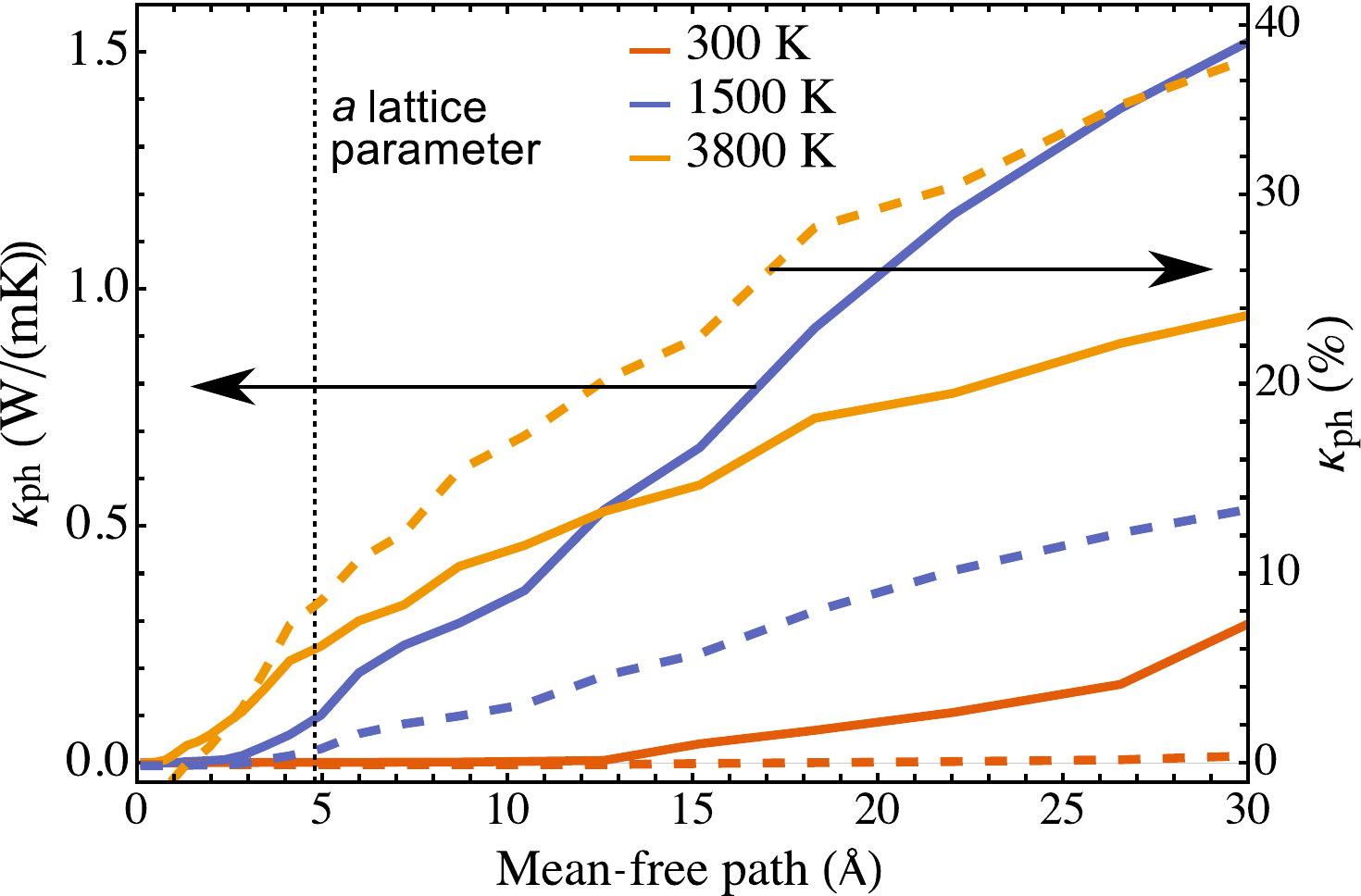}
\caption{Accumulated phonon thermal conductivity as a function of mean-free path, shown in solid lines, left axis. Percentage accumulated phonon thermal conductivity is shown in dashed lines, right axis. \label{fig: mfpEffects} }
\end{figure}

Temperature mean-free path saturation effects are expected to be weak until very high temperatures based on Fig. \ref{fig: mfpEffects}. At 300 K, $\kappa_\text{ph}$ from mean-free paths shorter than 10 Å is negligible. At 1500 K, less than 1\% of the computed $\kappa_\text{ph}$ arises from mean-free paths comparable to the lattice parameter. Even at 3800 K, when the scattering rate is extremely high for the material, more than 90\% of the computed $\kappa_\text{ph}$ comes from mean-free paths greater than $a$, with approximately $0.2$ $\text{Wm}^{-1}\text{K}^{-1}$ associated with mean-free paths shorter than the lattice parameter.

\section*{Phonon renormalization and four-phonon scattering\label{sec:Phonon-scattering}}

\begin{figure*}
\includegraphics[scale=0.7]{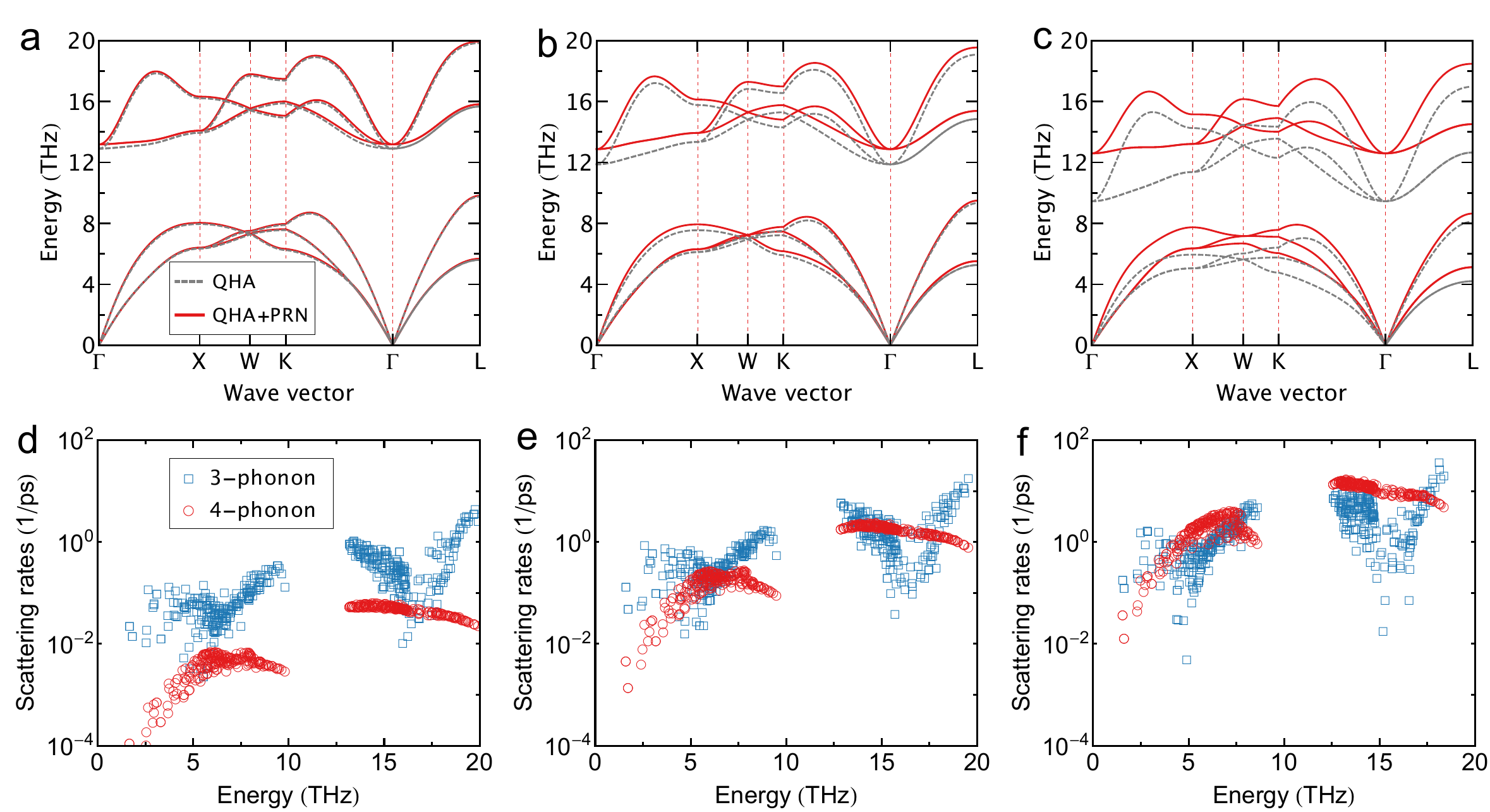}
\caption{Phonon dispersion at 300 K (a), 1500 K (b) and 3800 K (c) at the quasiharmonic and quasiharmonic+renormalization levels of theory. Three and four-phonon scattering at 300 K (d), 1500 K (e) and 3800 K (f). \label{fig: prn4ph} }
\end{figure*}

At high temperature quasiharmonic and anharmonic frequency renormalization, along with four-phonon scattering, are expected to play increasingly more important roles. Quasiharmonic volume expansion generally lowers phonon frequencies, as shown in Fig. \ref{fig: prn4ph}a-c, while anharmonic phonon renormalization tends to have the opposite effect and harden frequencies in ZrC. Compared to three-phonon scattering, four-phonon scattering is a minor effect at low temperature (300 K), but becomes very prominent for temperatures exceeding 1500 K as shown in Fig. \ref{fig: prn4ph}d-f.

\end{document}